\documentstyle[12pt,aaspp4]{article}

\def\kms{km\thinspace s$^{-1}$}
\def\cm2{cm$^{-2}$}
\font \sc = cmr10
\font \itsmall = cmti9
\font \bfsmall = cmbx9
\font \bfti = cmbx10
\def\HI{H{\sc I}}
\def\HII{H{\sc {II}}}

\def\HIIit{H{\itsmall{II}}}
\def\HIbf{H{\bfsmall{I}}}
\def\HIIbf{H{\bfsmall{II}}}
\def\HIbfti{H{\bfti{I}}}
\def\HIIbfti{H{\bfti{II}}}
\def\etal{{\it et al.\/\ }}

\def\eg{{\it e.g.,\/\ }}
\def\sun{_\odot}
\input psfig.tex

\begin{document}

\title{The \HIbfti\ Companions of \HIIbfti\ Galaxies and \linebreak
Low Surface Brightness Dwarf Galaxies}
\vspace{.5in}
\author{Christopher L. Taylor}
\vspace{.1in}
\affil{McMaster University \linebreak Department of Physics and Astronomy, 
\linebreak Hamilton, Ontario, Canada, L8S 4M1 \linebreak 
{\it taylorc@physun.physics.mcmaster.ca}}
\authoraddr{Department of Physics and Astronomy, Hamilton, Ontario,
Canada, L8S 4M1}

\vfill

{\it Note: This paper has been accepted for publication in The 
Astrophysical Journal.}

\vfill
\eject

\begin{abstract}

I study the VLA \HI\ survey of \HII\ galaxies by Taylor \etal (1995, ApJS, 99, 
427; 1996, ApJS, 102, 189) and the VLA \HI\ survey of low surface brightness
(LSB) dwarf galaxies by Taylor \etal (1996, in press) to investigate the role
of galaxy interactions in triggering the bursts of massive star formation
seen in \HII\ galaxies.  Comparing the two surveys, I find that \HII\
galaxies have companions more than twice as often as LSB dwarfs ($p = 0.57$
for \HII\ galaxies, compared to $p = 0.24$ for LSB dwarfs).  I examine
the completeness of the companion samples detected by the two surveys.  
For the companions to \HII\ galaxies, the sample is likely complete in the
distribution of velocity separations from their parent galaxies, but
is probably missing some companions at large projected linear separations
because of the finite size of the VLA primary beam.  For the companions of
LSB dwarfs, the small number of detections means their distributions in
velocity and linear separation are poorly determined, but the LSB dwarfs 
were observed with the same observational setup as the \HII\ galaxies so 
they will have the same levels of completeness.  Because the two samples 
were observed in exactly the same fashion, there will be no relative bias 
in the number of companions introduced in this way.  In addition, 
the redshift distributions of the two samples are very similar, so there 
will not be a distance related relative bias.  

Thus I conclude that the difference in the number of \HI\ rich companions
is genuine, and signifies a difference in the local, small scale environments
between the two types of galaxy.  I search through published galaxy catalogs 
to determine number of neighbors each galaxy has outside the area of the VLA 
observations.  At these large separations, the number of neighbors is the
same, within the errors, for the two types of galaxy.  The high rate of 
companion occurrence at low separations for \HII\ galaxies relative to 
LSB dwarfs supports the hypothesis that the bursts of star formation are 
triggered by galaxy interactions.

\end{abstract}

\section{Introduction}

The small scale environment of a galaxy can have a profound impact on
that galaxy's evolution.  Obviously mergers and tidal interactions can
alter a galaxy's morphology by changing the distribution of its stars
and gas.  Interactions are also linked to nuclear star burst events, 
enhanced levels of massive star formation in galactic disks, and the
formation of bars (\eg Bushouse \markcite{Bu87}1987; Kennicutt \etal 
\markcite{Ke87}1987; Noguchi \markcite{No87}1987).  All of these 
processes are clearly most efficient
in an environment of high galaxy density, where the likelihood for
galaxies to interact is greatest.  Even the type of galaxies available
as interaction partners can be important, as shown by the examples of
elliptical galaxies believed to have accreted gas from neighboring
late type systems (\eg NGC~1052; van Gorkom \etal \markcite{vG86}1986).

Taylor, Brinks \& Skillman \markcite{TBS}(1993; hereafter TBS) used the 
idea that such 
interactions might trigger star formation to launch a program of searching 
for previously unknown companions.  They selected a sample of nine dwarf 
galaxies currently experiencing a burst of star formation (\HII\ galaxies), 
but without obvious interaction partners.  If the bursts of star formation 
were related to interactions, then the companions must be optically faint
to have avoided earlier detection.  Dwarf galaxies were chosen because
they are too small to sustain spiral density waves, which can trigger star 
formation episodes in spiral galaxies.  TBS \markcite{TBS} observed these
\HII\ galaxies with the NRAO\footnotemark \footnotetext{The National 
Radio Astronomy Observatory is a facility of the National Science Foundation 
operated under cooperative agreement by Associated Universities, Inc.}
Very Large Array (VLA) in the 21 cm transition of \HI, looking for
optically faint but \HI\ rich companions.  They found four of the nine
had previously unknown companions.

Following the success of the pilot study of TBS, Taylor \etal 
\markcite{TBGS}(1996a, 1995; hereafter TBGS) conducted a similar survey 
around a larger (N = 21), volume limited sample of \HII\ galaxies, to put 
the result on a more solid statistical footing.  They detected companions 
around 12/21 ( = 0.57) \HII\ galaxies.  Taylor \etal \markcite{TTBS}(1996b, 
in press; hereafter TTBS) then surveyed a complementary sample (N = 17) of 
low surface brightness ({\it i.e.} non--starbursting) dwarf galaxies to serve 
as a control on the \HII\ galaxy sample.  TTBS found that only 4/17 ( = 
0.24) of the LSB dwarfs had \HI-rich companions.  Thus the \HII\ galaxies 
are more than twice as likely to have \HI\ rich companions as the LSB dwarfs.

The goal of this paper is to compare the data from these two \HI\ surveys,
to determine whether or not the presence of a nearby \HI-rich companion can
trigger bursts of star formation in low mass galaxies.  In Section~2 I 
review the properties of the \HII\ galaxy and LSB dwarf samples.  
In Section~3, 
in which I compare the completeness of the companion searches for the two 
surveys and discuss their observational biases.  I will show that the 
difference in the observed numbers of companions is not caused by such 
effects.  In Section~4 I compare the large scale (0.5 -- 2.5 Mpc) environments 
of the two samples, showing that on average, on the {\it large scales}, the 
number of neighbors around the \HII\ and LSB galaxies is the same. Section~5 
is a discussion of the relationship between interactions and star formation 
in dwarf galaxies.  Section~6 presents the conclusions. 

Throughout this paper I adopt the following definitions: a companion
to one of the \HII\ galaxies or LSB dwarfs means one of the objects detected
in the VLA \HI\ surveys described above, while a neighbor is a cataloged 
optical galaxy in the vicinity of one of the \HII\ galaxies or LSB dwarfs.  
Note that there is a degree of overlap between these two classifications, 
as some of the companions are cataloged galaxies (\eg UM477A = NGC4116).

\section{A Description of the Two Samples of Galaxies}

The search for companions to \HII\ galaxies is described in detail by TBGS.
The sample of \HII\ galaxies was selected from the galaxies studied by 
Salzer\markcite{SMB1}\markcite{SMB2} \etal (1989a,b).  Because they 
originally come from an objective prism survey, these galaxies tend to have 
emission lines that are strong relative to their stellar continuum.  Thus 
galaxies with inherently strong emission lines {\it or} a faint continuum 
are selected.  The latter case can result in the inclusion of extremely
low mass systems, while the former can include AGNs.  In order to restrict
the sample to relatively nearby dwarf galaxies, TBGS applied a velocity 
limit of $v\sun \leq$ 2500 \kms\thinspace and an absolute magnitude limit
 of $M_B \geq -19$.
Despite these selection criteria, two non-dwarfs -- low luminosity spirals,
were included in the sample.  I will discuss the effects of these two 
galaxies on the results of this paper in Section~5.2.  The \HI\ masses 
of the \HII\ galaxies range from a few $\times 10^7$ M$\sun$ to a few 
$\times 10^9$ M$\sun$.  TBGS found that 12/21 of those \HII\ galaxies had
nearby companions rich in \HI, for a measured companion frequency of 0.57, 
with a strong lower limit of 0.37. Because of the observational limitations 
(discussed in TBGS) it was not possible to derive a useful upper limit.

TTBS provide a detailed account of the search for companions to LSB dwarfs.
The LSB dwarf galaxies were extracted from the list of low surface brightness
galaxies studied by Bothun \etal \markcite{BSISM}(1993).  To be included in 
the list used by Bothun \etal, \markcite{BSISM} a galaxy had to have a low
central surface brightness and exceed a minimum radius.  The lists of LSB 
galaxies are not statistically complete samples (\eg Schombert \& Bothun 
\markcite{SB88} 1988) and thus it is unknown how representative these 
galaxies are of whole LSB galaxy population.  To obtain a sample useful for 
comparison with the \HII\ galaxies, TTBS selected LSB dwarfs with a similar 
velocity limit (3000 \kms) and to resemble the \HII\ galaxies in morphology, 
\HI\ mass and \HI\ linewidth.  For example, the range in \HI\ mass of the
LSB dwarf galaxies is few $\times 10^7$ M$\sun$ to a few $\times 10^9$ 
M$\sun$, very similar to that of the \HII\ galaxies.  TTBS performed a 
Kolmogorov--Smirnov (K--S) test on the distribution of \HI\ masses for the 
two samples and found the result consistent with the two samples coming from
the same parent population. In the search for companions, TTBS found that 
only 4/17 ( = 0.24) had \HI-rich companions.  Here again an upper limit was
problematic.

It is important to distinguish between LSB and dwarf galaxies.  The 
description ``low surface brightness'' generally refers to the central
surface brightness of a galaxy being below some limit usually set at
about 23 magnitudes arcsec$^{-2}$.  Dwarf galaxies often meet this 
criterion, although there are large numbers of LSB spiral galaxies 
(\eg Schombert \etal \markcite{SBSM}1992).  As described above, TTBS
used selection criteria to obtain a sample of galaxies that are {\it both} 
LSB {\it and} dwarf, and the similarities in \HI\ properties between their
LSB dwarf sample and the \HII\ galaxy sample shows they were successful.

The samples of \HII\ galaxies and LSB dwarfs were selected to be as similar
as possible, within the constraints of the source catalogs from which they
were drawn.  In addition, identical observational parameters were used in 
conducting each survey.  These similarities reduce the prospect of one 
sample being biased toward having more companions relative to the 
other.  As an example of the similarity between the two samples, Figure~1 
shows histograms comparing their velocity distributions.  Clearly the two 
are very similar, and the K--S test shows that the probability that the two 
distributions come from the same parent population is 0.98.  In the next 
section I will examine the sources of a possible relative bias between the 
two samples and show that such a bias does not exist.
\placefigure{fig1}

\section{The Completeness of the Companion Samples}

\subsection{The Completeness in Radial Velocity Separations}

The necessity of a finite number of channels and the desire for reasonable 
velocity resolution restrict  the region of velocity space around each 
target galaxy that can be included in the \HI\ surveys.  This could lead 
to a bias in the observed distribution of radial velocity differences by 
excluding companions with high velocities relative to their parent galaxies.  
I will examine this distribution for the companions of both the \HII\ 
galaxies and the LSB dwarf galaxies to investigate the possibility of this 
bias.

\subsubsection{The Companions of the \HIIit\ Galaxies}

The upper panel of Figure~2 shows a histogram of the absolute value of radial 
velocity separations for the \HII\ galaxy companions (solid line).  The 
velocity coverage of the VLA observations is $\pm$ 250 \kms.  Not a single 
\HII\ galaxy / companion 
system has a separation greater than 110 \kms, despite the fact that the 
observable range is more than twice that.  The dot-dash line shows the 
distribution for a random sample of velocity separations compiled by 
shuffling the radial velocities of all the \HII\ galaxies and companions 
and forming random pairs (Turner \markcite{T76b}1976{\it b}).  This 
represents what would be seen if there were no true physical pairs and 
only random alignments in velocity space.  The distribution of velocities 
towards low values is also consistent with a sample of true physical 
pairs, as false detections of companions resulting from noise spikes would 
be distributed randomly within the velocity coverage of the observations,
not clustered toward low velocities.
\placefigure{fig2}

The observed distribution falls off dramatically and approaches the limit 
of the random distribution by 110 \kms.  Studies of velocity distributions
of paired galaxies been done by several authors for samples of binary 
galaxies drawn from catalogs which consist largely of ``normal'' galaxies 
({\it i.e.} M$_B \leq$ $-$19).  Turner (\markcite{t76a}1976{\it a}) found 
a cutoff for physically associated pairs at $\Delta v = $ 450 \kms, whereas 
Peterson \markcite{P79}(1979) adopted a value of 750 \kms, and van Moorsel 
\markcite{vM87}(1987) used 500 \kms.  The difference between the 110 \kms\ 
cutoff of the TBGS sample and the much large cutoff for the binary spiral 
samples reflects the differing mean mass of the \HII\ galaxy sample compared 
to the ``normal'' galaxy samples.  The escape velocity scales as the 
one-half power of the parent galaxy mass.  The ratio of mean mass between 
the TBGS sample and that of Peterson \markcite{P79} is 40.8.  Scaling 110 
\kms\ up by (40.8)$^{\slantfrac{1}{2}}$ gives 700 \kms, very close to 
the cutoff velocity separation found by Peterson. \markcite{P79}  Galaxies 
with more mass are able to retain companions with higher relative velocities. 

Because the distribution of velocity separations falls off well before
the end of the range of velocity coverage of the VLA, and because these
counts seem to merge smoothly into the expected distribution of random
background separations, I conclude that the sample of companions to \HII\ 
galaxies is likely complete in velocity space.  

\subsubsection{The Companions of the LSB Dwarf Galaxies}

Although the LSB dwarf galaxies have far fewer companions than the \HII\
galaxies, the distribution of radial velocity separations shown in the 
lower panel of Figure~2 (solid line) does resemble the distribution of the 
\HII\ galaxy companions in general shape.  That is, the distribution peaks 
at low velocities and decreases at higher velocities.  Unfortunately the 
small number of companions found around the LSB dwarfs makes it difficult 
to argue that distribution is truly similar to the distribution of \HII\
galaxy companions.  Like the \HII\ galaxy companions, the largest velocity 
separation of the LSB dwarf companions is much lower than the maximum 
observable separation of 250 \kms, but again, this could just be a result
of the small number of detections.  If the LSB dwarfs exist in an 
environment similar to the environments of the \HII\ galaxies, the 
distributions of companions at large velocity differences will be similar 
as well.  However, TTBS found that most of the companions to the LSB dwarfs
were more massive galaxies, unlike the case for the \HII\ galaxies, where
the companions were of equal or less mass.  Therefore, because the LSB dwarfs
are bound to more massive galaxies, their velocity separations could be 
larger than was true for the \HII\ galaxies while still allowing the systems
to be bound.

There is no evidence in the radial velocity distribution of companions to 
LSB dwarfs that there exist companions at large velocity separations.  
I conclude that I have not missed a large number of companions within
one primary beam of the target galaxies due to the finite velocity coverage
of the TTBS VLA observations.

\subsection{The Completeness in Projected Physical Separation}

The problem of assessing the completeness of the companion sample in spatial 
separation is entirely different from the problem of velocity separation. 
This is because the amount of velocity space observed by the VLA does not 
depend upon redshift, whereas for a constant angular size of the VLA primary 
beam, a larger physical area is imaged as the distance increases.  It is 
immediately apparent that companions at relatively large projected
separations may fall outside the primary beam for galaxies at smaller
distances.

\subsubsection{The Companions to the \HIIit\ Galaxies}

The ratio of the largest and smallest distances in the \HII\ galaxy 
sample is 2.6, corresponding to a ratio of imaged areas of 6.8.  The 
distribution of projected separations for the companion population of 
\HII\ galaxies is shown in the upper panel of Figure~3.  The vertical line 
in the Figure indicates the radius of the FWHM of the VLA primary beam 
for UM533, the closest \HII\ galaxy.  At least one of the distant \HII\ 
galaxies has a companion beyond this radius, suggesting that some of the 
nearer galaxies could as well. Such companions would have been undetected 
by TBGS. \markcite{TBGS}
\placefigure{fig3}

An additional effect adding to the incompleteness of the survey is the
decrease in sensitivity of the VLA primary beam with distance from the
pointing center.  For observations in the 21--cm line, the sensitivity
falls to half maximum at a radius of 15\arcmin.  The farther from the
target galaxy a companion lies, the more \HI\ it must have to be
detectable.  Figure~4 shows the distribution of angular separation for
the companion population.  The dashed vertical line shows the radius
of the FWHM of the VLA primary beam.  The mean \HI\ mass of the four
systems beyond the line is 18.5 $\pm$ 4.4 $\times$ 10$^8$  M$\sun$, while 
for the entire population it is 6.5 $\pm$ 2.7 $\times$ 10$^8$  M$\sun$,
illustrating the effect of the decreasing sensitivity.  For the companion
with the widest angular separation in Figure~4 the decrease in sensitivity
is approximately 0.17.  A typical sensitivity for the distances of the
\HII\ galaxies is $\sim$ 3~$\times$~10$^7$ M$\sun$ (discussed in greater
detail in the following section).  For the widest separation, this limit
becomes $\sim$ 2~$\times$~10$^8$ M$\sun$.  Five companions have low enough
\HI\ masses to become undetectable at this separation, and even six of the
\HII\ galaxies would have been missed if they had been towards the edge
of the primary beam.  Clearly this effect could result in companions
being missed in the survey.
\placefigure{fig4}

Although it is likely that companions have been missed at large radii
for the reasons discussed above, it is also possible that the true
distribution of separations will tend towards companions with low 
separations, if the companions experience dynamical friction from dark 
matter halos in the \HII\ galaxies (Lin \& Tremaine \markcite{LT83}1983).  
For typical projected 
separations ($r =$ 60 kpc) and radial velocities ($v =$ 40 \kms), the 
crossing time is of order 6~$\times$~10$^9$ yr.  If the dark matter 
halos reach the distance of the companions and the companions are on
circular orbits, then the timescale for dynamical friction can be written
as:
$$ t_{fric} = \frac{1 \times 10^{10}}{\ln \Lambda} \left(\frac{r}{60\ {\rm
kpc}}\right)^2 \left(\frac{v}{220\ {\rm km/s}}\right) 
\left(\frac{2 \times 10^{10} {\rm M\sun}}{M_{HII}}\right) \ {\rm yr} $$

where

$$\Lambda = \frac{r v^2}{G(M_{HII} + M_{comp})} $$

(Binney \& Tremaine \markcite{BT87}1987).

For $M_{HII}$ = 10$^9$ M$\sun$ and $M_{comp}$ = 10$^8$ M$\sun$ this
timescale is approximately 1.2~$\times$~10$^{10}$ yr.  Thus the companions
would have had plenty of time to experience dynamical friction, but not
enough to have been swallowed by the \HII\ galaxies yet.  This time scale
is, of course, lower limit because I have used observed orbital parameters 
like projected separations and radial velocities for the true separations 
and relative velocities. 

The conclusion to be drawn regarding the completeness of the companion 
sample in physical separation is a fairly weak one.  Because of the range
in distance of the \HII\ galaxies, and the fall off of sensitivity at the
edges of the VLA primary beam, it is very likely that some companions have 
been missed. Unfortunately, because the distribution of the companions is 
unknown, the degree of incompleteness cannot be estimated. 

\subsubsection{The Companions of the LSB Dwarf Galaxies}

The ratio of the largest and smallest distances in the LSB dwarf galaxy 
sample is 2.5, corresponding to a ratio of areas imaged of 6.3, nearly 
identical to the values for the \HII\ galaxies.  This is because the
LSB dwarfs were chosen to span approximately the same range in redshift
as the \HII\ galaxies (recall Figure~1).  

The distribution of projected separations for the companion population of 
LSB dwarfs is shown in the lower panel of Figure~3.  The vertical line 
indicates the radius of the FWHM of the VLA primary beam at the distance 
of the closest galaxy in the LSB sample.  The companions of the LSB dwarfs 
are, on average, further away from the LSBs than the companions of \HII\ 
galaxies are from the \HII\ galaxies.  This may, however, be a function of 
the small number of detected companions.  It is also possible that because 
the companions to LSBs are mostly more massive galaxies the LSB/companion 
systems can remain bound to larger distances for roughly equivalent 
velocity separations.

Even less can be said about the completeness in spatial separation of the 
LSB dwarf sample than could be  about that of the \HII\ galaxies.  If it is 
true that the companions of LSB dwarfs tend to larger physical separations, 
then the LSB dwarf sample is likely {\it more} incomplete in the companion 
count than is the the \HII\ galaxy sample.  In this case, the variation
of sensitivity of the VLA primary beam with radius will affect the LSB
dwarf sample more than it does the \HII\ galaxy sample.   However, if 
the distribution seen in Figure~3 is just the result of a small sample size, 
then the level of incompleteness is likely the same as for the \HII\ galaxies,
in which case the primary beam effect will be the same on both samples.  
Thus {\it either\ }:  
\begin{enumerate}

\item the companions to LSB dwarfs are distributed at larger linear 
separations than are the companions of \HII\ galaxies, in which case 
the incompleteness in the companion counts is greater than for the 
\HII\ galaxies; {\it or} 

\item the two distributions are roughly similar, in which case the 
incompleteness of the two samples is roughly the same.

\end{enumerate} 

Based on the current data there is no way of determining which of these
possibilities is correct.

\subsection{Sensitivity Limits to Detectable Masses of \HIbf\ Companions}

Another way in which companions might be missed by TBGS \markcite{TBGS} 
or TTBS \markcite{TTBS} is if they do not have enough \HI\ to be detected.
This is especially acute for dwarf ellipticals, as dEs have very little 
or no \HI, but it will also affect the detection of low mass dwarf 
irregulars (see the discussion below).

\subsubsection{The Companions of the \HIIit\ Galaxies}

Regarding dwarf elliptical as companions, I simply note that only one 
candidate was seen in any of the optical data (see UM465 in TBGS), although 
the area of the CCD on the sky was much smaller than the area of the 
\HI\ data.  I will discuss a search for optical companions performed
using the Center for Astrophysics (CfA) redshift survey in Section~4.

The upper panel of Figure 5 shows \HI\ masses for the sample of {\it both} 
\HII\ galaxies 
and companions, plotted as a function of redshift (distance) .  The solid 
line shows the sensitivity of the TBGS \markcite{TBGS} observations with 
distance for an unresolved point source with a typical single channel 
noise of 1.3 mJy beam$^{-1}$.  The lower envelope of \HI\ mass of the 
combined sample is roughly constant with distance, at a few times 
10$^7$ M$\sun$, even at small distances where the detection threshold 
is significantly lower.  The only point below this constant level 
corresponds to the \HII\ galaxy UM~538, at a distance of 14.3 Mpc.  It 
is the only galaxy {\it or} companion that could possibly have been 
lost if it were at a higher redshift within the limits of the survey.  
I take this as an indication that there are not large numbers of 
companions of \HI\ mass $\sim$ 10$^7$ M$\sun$ missing from the sample at 
the larger distances.  This is a result of the relatively narrow range 
in velocity covered by the \HII\ galaxies (from 700 \kms\ to 2500 \kms).  
On the other hand, there are a number of dwarf irregular galaxies in and
around the Local Group with \HI\ masses in the range $\sim$ 10$^6$ to
10$^7$ M$\sun$ (\eg Leo~A, Sag DIG, GR~8; Lo, Sargent \& Young 
\markcite{LSY}1993).  Such objects are not limited to the Local Group
(C\^{o}t\'{e} 1995\markcite{C95}), so there could be a significant 
population of companions at low 
\HI\ mass (of order 10$^6$ M$\sun$ or less) which would remain 
undetected unless closer \HII\ galaxies were surveyed, or else a factor 
of 100 more integration time is spent on the current sample.  
\placefigure{fig5}

\subsubsection{The Companions of the LSB Dwarf Galaxies}

TTBS obtained optical images of each of the LSB dwarfs using the STScI 
Digitized Sky Survey.  None of the systems had a potential dwarf elliptical 
companion visible in the images, though the images were restricted to a 
small region around each galaxy ($\sim$ 7.5$\arcmin$).

A plot of \HI\ mass versus distance for the LSB dwarfs and their companions
is shown in the lower panel of Figure~5.  The detection limit is nearly 
identical to the 
limit for the \HII\ galaxy observations, being slightly lower because most
of the LSB dwarf observations took place after sunset, whereas nearly all
of the \HII\ galaxy took place during the day and close to the solar maximum.
Thus the LSB dwarf observations are less affected by solar interference,
although they are not unaffected (TTBS).

As was true for the \HII\ galaxies and their companions, the detections of 
LSB dwarfs and their companions do not approach the detection limit, 
suggesting that there are not large numbers of objects hiding below the 
detection threshold.

\subsection{Summary of the Two Samples' Completeness}

In the above subsections it was shown that the \HII\ galaxy companion
sample is likely complete in the distribution of radial velocity separations.
Although there are not enough detections of companions around LSB dwarfs 
to show this is also true for that sample, the similarity of both the target 
galaxies' properties and the observing conditions makes it highly likely.   
It was also shown 
that the \HII\ galaxy companion sample is likely incomplete in projected
linear separation, in the sense of missing companions around the most nearby
\HII\ galaxies at separations greater than $\sim$ 100 Mpc.  Because the
LSB dwarfs have approximately the same distribution in redshift, this is
likely true for them as well.  Finally it was argued that the two samples
are not likely to be missing large numbers of companions at the sensitivity
level of the observations ($\sim$ 10$^7$ M$\sun$), although the data do not
constrain the number of less massive companions.  In any event, the sensitivity 
of the two surveys to \HI\ mass is nearly identical, so the effects of this
sensitivity limit will be the same on each sample.

The most important fact to keep in mind for this work is the great 
similarity between the two surveys being studied, in terms of the
redshift distributions, galaxy masses and observing setups.  Because of
this, whatever biases imposed due to limitations in the samples 
(\eg the projected linear separation) will affect both the \HII\ galaxy and
LSB dwarf samples equally.  Therefore the difference between the companion
rates of the two types of galaxy noted in Section~1 and by TTBS 
\markcite{TTBS} is real, and not caused by a relative bias between the 
samples.

\section{A Comparison of the Large Scale Environments of \HIIbf\ and
Low Surface Brightness Dwarf galaxies}   

The previous section showed that the difference in the number of \HI\
companions around LSB dwarf and \HII\ galaxies is not due to selection
effects effects inherent in galaxy samples, nor is it due to limitations
imposed by the observing conditions or setup.
The observed difference in the number of companions therefore reflects a
true difference in the small scale environment of the two populations.
However, it is likely that the small scale environment is influenced
by the larger scale environment (\eg cluster versus field, or supercluster
versus void).  If, to give an extreme example, all the \HII\ galaxies 
were found to be in the center of the Virgo Cluster, while all the LSB 
dwarfs were on the edge of a void, then the observation that \HII\ galaxies 
have more companions would be a reflection of the generally denser environment
of the particular sample of \HII\ galaxies and not necessarily indicating
an important difference in their evolution.  On the other hand, if the 
two types of galaxy were found to have the same large scale environment,
then the difference in the companion rates would be significant.

To investigate the large scale environments of the two samples in I will 
compare them to two galaxy catalogs: the CfA redshift survey catalog
(Huchra \etal \markcite{HDLT}1983) and the Nearby Galaxies Catalog 
(Tully \markcite{Tu88}1988). 

\subsection{Comparison to the CfA Redshift Survey}

The area of overlap of the CfA redshift survey with the TBGS and TTBS is in
two regions, $b_{II} \geq 40\arcdeg$, $\delta \geq 0\arcdeg$ and
$b_{II} \leq -30\arcdeg$, $\delta \geq 2\arcdeg.5$.  In these areas the
catalog is complete to $B \leq 14.5$ (Huchra \etal \markcite{HDLT}1983).
For the most and least distant galaxies in the two samples, this 
corresponds to $M \leq -18.4$ and $M \leq -15.5$ respectively.  Only for
the closest galaxies will any of the neighbors be dwarfs.  A comparison
with primarily massive galaxies will still be useful, because dwarf galaxies
are known to trace the same structures as bright galaxies (\eg Thuan \etal
\markcite{TAGS}1991)

These areas contain 11/21 of the \HII\ galaxies, and 15/17 of the LSB
dwarfs.  I obtained the catalog from the Astrophysics Data Service (ADS) 
Catalog Query Service available on the World Wide Web.  I searched 
through the catalog looking for neighbors around each galaxy with 
velocities within 500 \kms\ of the target galaxy.  The choice of 500 \kms\
as the velocity limit was not based on any physical reasons, but rather
to allow for easy comparisons with similar studies by Bothun \etal 
\markcite{BSISM}(1993) and Campos--Aguilar \& Moles \markcite{CM}(1991) which
adopted that limit.  Note that 500 \kms\ is consistent with the typical
velocity limits for samples of binary spirals discussed in section~2.1.

The counts were binned in 0.5 Mpc bins running from 0 to 2.5 Mpc.  Figure~6 
shows the mean neighbor counts plotted versus distance for the \HII\ 
galaxy (triangles) and LSB dwarf galaxy (circles), with associated error 
bars.  The open symbols show the counts in each bin, while the filled points
show the cumulative values.  In all bins the counts per bin are the same 
within the errors, although in the last two bins 
(2 and 2.5 Mpc) the LSB dwarf neighbor counts begin to diverge from the 
\HII\ galaxy neighbor counts.  It is possible that if the samples were 
larger (and thus the errors smaller) that there might be a significant 
difference.  The counts at smaller radii, however, are nearly identical, 
suggesting that the members of the two samples are in environments of 
similar galaxy density.  From neighbor counts using the CfA redshift survey
catalog, I conclude that, on average, the \HII\ galaxies and LSB dwarfs are
found in environments of similar galaxy density on scales larger than
the VLA primary beam ({\it i.e.,} $\sim$ 250 kpc).
\placefigure{fig6}

How does this result compare with other studies of galaxy environments?
Bothun \etal \markcite{BSISM}(1993) have done a very similar analysis on
much larger samples (N between 135 and 870) for galaxies in the redshift 
range 2000 -- 12000 \kms.  Most of the galaxies from that work are not 
dwarfs like the galaxies of this study, but rather are more massive 
spirals and ellipticals, with large \HI\ line widths.
For the portion of the Bothun \etal \markcite{BSISM}(1993) sample cataloged
by Schombert \etal \markcite{SBSM}(1992), the mean \HI\ line width measured
at half the maximum intensity is 147 $\pm$ 8 \kms, while for the LSB dwarf 
sample of TTBS\markcite{TTBS} the line width measured at zero intensity is
only 107 $\pm$ 8 \kms.  If the LSB dwarfs had been measured on the 
half--maximum too the line widths would be even smaller, emphasizing even
more the difference between the Bothun \etal \markcite{BSISM} and the
TTBS LSB samples.

Bothun \etal \markcite{BSISM}(1993) searched for neighbors within a 
projected radius of 2.4 Mpc and $\pm$ 500 \kms\ of each of their sample 
galaxies. They binned their neighbor counts in 0.5 Mpc bins, finding
that at every bin the cumulative number of neighbors for HSB galaxies
was larger than for LSB galaxies. The difference was most pronounced
for the 0 -- 0.5 Mpc bin.   This is consistent with my VLA result 
that in the range of distances 0 -- 0.25 Mpc \HII\ galaxies have more
\HI\ companions than do LSB dwarfs.  In my neighbor counts using the CfA
redshift survey, however, I find that the 
\HII\ galaxies and the LSB dwarfs have the same number of neighbors at 
{\it all} separations within the errors, although the counts do seem to
be diverging from each other at larger radii.  Thus based only on the 
neighbor counts from optical catalogs, my results would not agree with 
the Bothun \etal result, but this may simply be because I have far fewer 
galaxies in my samples than they had, resulting in larger error bars.  The 
Bothun \etal samples have from 7 to 50 times as many galaxy as the samples 
of this paper. When including the VLA discovered companions, I find the \HII\ 
galaxies have an excess of companions in the 0 -- 0.25 Mpc range, but most 
of these companions are not bright enough to be in the catalogs searched 
by Bothun \etal \markcite{BSISM}(1993).  Thus within the errors of the
neighbor counts, there is no significant difference in the environments of 
the two samples on the scales of 0.5 to 2.5 Mpc.  Overall, I find that
LSB dwarfs have fewer companions at short radii than do \HII\ galaxies,
but at larger radii the two sample have similar numbers of (optical)
neighbors.  This is the same result arrived at by Bothun \etal 
\markcite{BSISM} for their samples of LSB and HSB galaxies.

Other studies have investigated the environments of \HII\ galaxies, but 
unlike my study and that of Bothun \etal \markcite{BSISM}(1993), they did 
not compare HSB and LSB systems.  Work by Campos--Aguilar \& Moles 
\markcite{CM}(1991), Campos--Aguilar \markcite{CMM}\etal (1993) and Telles 
\& Terlevich \markcite{TT}(1995) has found that \HII\ galaxies tend to 
be isolated from luminous, massive galaxies. These authors
used various optical galaxy catalogs to search for bright neighbors.
Typically from one-third to three-fourths of the \HII\ galaxies were
found to be isolated, with isolation criteria of 1 Mpc and $\pm$ 500
\kms.  By restricting themselves to magnitude limited optical catalogs,
however, these authors could not have found the faint dwarfs that form
the majority of the optical counterparts to the TBGS \markcite{TBGS}
\HI\ detections. 
 
For a quantitative comparison, 19 out of 21 \HII\ galaxies from the TBGS
\markcite{TBGS} sample are also included in the Campos--Aguilar \& Moles
\markcite{CM}(1991) sample.  If we consider these 19 \HII\ galaxies then
using just the Campos--Aguilar \& Moles \markcite{CM}(1991) detections of
neighbors, the fraction of isolated galaxies is 0.11.  The fraction of 
isolated \HII\ galaxies for their entire sample is much higher, 0.33.  With
such a high fraction of isolated \HII\ galaxies in their sample they argue 
that galaxy interactions are not responsible for the bursts in {\it all} 
\HII\ galaxies.

This difference between the 0.11 and 0.33 fractions occurs because the TBGS
\markcite{TBGS} sample has been restricted to less than 2,500 \kms.  All
these galaxies lie within the Local Supercluster, whereas the much larger
sample of Campos--Aguilar \& Moles \markcite{CM}(1991) contains galaxies of
velocity up to $\sim$ 15,000 \kms\thinspace and in a variety of large scale
environments. In fact, Campos--Aguilar \markcite{CMM}\etal (1993) do find
that the mean redshift of \HII\ galaxies without neighbors is higher than 
that of those with companions.  They suggest this is because the CfA redshift
survey, which they use to search for companions, is magnitude limited,
making faint neighbors at higher redshifts more difficult to detect.  It is
likely, therefore, that a VLA \HI\ survey of the entire Campos--Aguilar \&
Moles \markcite{CM}(1991) sample would yield a higher fraction of galaxies
with interaction partners than is currently known.

\subsection{Comparison to the Nearby Galaxies Catalog}

As mentioned previously, the CfA survey does not have complete coverage
over the area of all the galaxies in the \HII\ galaxy and LSB dwarf samples.  
Because of this I could only compare the parts of those samples that
did overlap with the CfA coverage.  The sample in the Nearby Galaxies
Catalog (hereafter NBG; Tully \markcite{Tu88}1988), however, has nearly uniform
coverage across the entire sky (except what is obscured by the Galactic
plane).  This catalog has roughly the same number of galaxies as the
portion of the CfA survey used above, but distributed across the entire sky.  
Therefore the limiting magnitude is not as deep.  Rather than compute 
projected separations with the somewhat sparsely distributed galaxies in 
the NBG, I will compare the positions and velocities of the sample galaxies 
with the positions and velocities of the groups defined in the Catalog.  
For this purpose I defined a sample from the Catalog that contained all
galaxies within three degrees and 500 \kms\ of a galaxy in either the
\HII\ galaxy or LSB dwarf samples.

An \HII\ galaxy or LSB dwarf was considered to be a group member if it
met both the following criteria: 
\begin{enumerate}

\item{} Its projected separation from the group center had to be less 
than twice the mean of the separations of each pair formed by pairing
each group member with every other group member, and; 
\item{} Its velocity had to be within two sigma
of the group mean velocity, where one sigma was the velocity dispersion of
the group.  

\end{enumerate}

This is a conservative definition, but one which is unlikely falsely call 
one of the sample galaxies a group member.  To test how well this definition 
works, I applied it to the NBG sample defined above, to see whether or 
not known group
members would be wrongly excluded.  Out of 32 groups with 131 members, only 
two galaxies (1.5\%) would have been excluded by my criteria.  Thus it is 
unlikely that I have incorrectly labeled an \HII\ or an LSB galaxy as not 
being a member of any group.  However, if incorrect assignments occur, 
they should happen with equal frequency to both samples, {\it assuming} the 
groups near the different samples are similar to each other.  To check the 
validity of this assumption, I calculated the separation of each pair in 
every group in the NBG sample.  The mean of this 
pairwise separation within groups near \HII\ galaxies is 0.48$\pm$0.10 Mpc, 
and 0.50$\pm$0.05 Mpc for groups near LSB dwarfs.  There is no evidence 
groups near \HII\ galaxies are different than groups near LSB dwarfs.

Using the above conservative definition of group membership, I searched
through the groups in the NBG sample, checking for any to which an \HII\
galaxy or LSB dwarf might belong.  I find that 5 out of 21 (= 0.24) \HII\ 
galaxies belonged to groups, while 7 of 17 (= 0.41) of the LSB dwarfs did.
Following TBS \markcite{TBS} and TBGS \markcite{TBGS} I can use the binomial 
distribution to calculate upper and lower limits on these fractions.  Using
this method to determine, for example, the lower limit for the number of
\HII\ galaxies in groups, will yield the lowest fraction of \HII\ galaxies
in groups which is consistent with observing 5 out of 21 at least five 
percent of the time if the experiment were repeated many times.  
With these calculations, the fraction of \HII\ galaxies in groups
is 0.24$^{+0.20}_{-0.14}$ and the fraction of LSB dwarfs in groups is
0.41$^{+0.23}_{-0.20}$.  These fractions are within the uncertainties of 
each other, indicating that there is no significant difference between the
group environments of the LSB dwarfs and the \HII\ galaxies.

\subsection{Contamination from the Virgo Cluster}

Nineteen of the twenty--one \HII\ galaxies from TBGS \markcite{TBGS} lie in
an area of the sky on the periphery of the Virgo Cluster.  All of these
nineteen are within 30$\arcdeg$\ of the cluster center.  Thus there is
the possibility that some of the \HII\ galaxy -- companion pairs are not
physically associated, but caused by a chance occurrence between a cluster
member and a nonmember at similar velocities.

To assess this possibility I will use the data of the Virgo Cluster study 
by Binggeli, Tammann \& Sandage\markcite{BTS87} (1987; hereafter BTS).
According to BTS\markcite{BTS87}, the Virgo Cluster contains several distinct
substructures.  Cluster A, centered on M87, and Cluster B, centered on M49 
are the two largest units, together containing most of the galaxies.  The
surface density of number counts for late--type galaxies can be described 
with an exponential profile: $I(r) = I_0 \exp(r/r_0)$, where $I_0 = 5$ 
arcsec$^{-2}$ and $r_0$ = 3.3 degrees.  Taking $r$ to be the angular distance 
of each \HII\ galaxy from the center of Cluster A (12$^h$ 25$^m$, 13$\arcdeg$ 
0$\arcmin$), this expression yields the number of Virgo galaxies 
expected within a VLA primary beam which is pointed at each \HII\ galaxy.
I then determine the fraction of Virgo galaxies which have velocities
within $\pm$ 250 \kms\ of each \HII\ galaxy to estimate the probability 
that if a galaxy did fall within the VLA primary beam, it would also be in
the correct velocity range so that TBGS would detect it.  Thus I obtain for
each observation of an \HII\ galaxy the number of members of Cluster A 
expected to be detected.  Summing this over all the \HII\ galaxies, only
0.16 galaxies are expected to contaminate the experiment.  Thus it is likely 
that no members of Cluster A are contaminating the experiment.

BTS\markcite{BTS87} did not derive a radial profile from the number counts
of Cluster B, presumably because it is not symmetric.  However, it lies
closer (12$^h$ 25$^m$.8, 8$\arcdeg$ 51$\arcmin$) to the TBGS \HII\ galaxies 
than does Cluster A, so some measure of the possible contamination is 
desirable. BTS\markcite{BTS87} found that Cluster B has one--fifth the 
population of Cluster A.  To get a rough idea of the possible contamination
caused by Cluster B, I will assume an exponential distribution as per Cluster
A, with the same scale length, but total number of galaxies scaled down by
a factor of five.  Repeating the calculation done for Cluster
A, shows that the number of contaminating galaxies is 0.09, less than
one, and less than what is expected from Cluster A.  I therefore conclude 
that interlopers from the Virgo Cluster are not likely affecting the counts
of companions.

Only eight of sixteen LSB dwarf galaxies from TTBS \markcite{TTBS} are 
within 30$\arcdeg$\ of the cluster center, but some of these LSB dwarfs
are closer to the center than any of the \HII\ galaxies.  Repeating the 
calculations above for the LSB dwarfs, I find the number of contaminating
galaxies to be expected is 0.41.  This is somewhat larger than for the 
\HII\ galaxies, but still so small as to make it unlikely.  However, if one 
of the \HI\ companions found by TTBS is in fact a chance alignment and not 
a physical pair, then removing it from the sample would make the companion 
rate even lower than 0.24 and increase difference between \HII\ galaxies and 
LSB dwarfs.

\section{Discussion}

\subsection{Interaction Triggered Star Formation in Dwarf Galaxies}

By comparing the data from TBGS \markcite{TBGS} and TTBS \markcite{TTBS}
I have shown that \HII\ galaxies have more \HI--rich companions than do
LSB dwarfs.  The fraction of \HII\ galaxies found to have such companions
is 0.57, with a lower limit of 0.37 (TBGS).  The fraction of LSB dwarfs
that have companions is 0.24, with a lower limit of 0.08 (TTBS).  In the
previous sections I showed that this difference is real, not caused by
observational biases, selection effects, or differences in the large scale
environments of the two samples.

Because \HII\ galaxies are currently experiencing bursts of massive star 
formation while LSB dwarfs have relatively low star formation rates, I 
interprete the difference in companion rates between the two types as 
evidence that interactions with such companions can trigger bursts of 
star formation.  This interpretation suggests that \HII\ galaxies may be 
LSB dwarfs in which interactions have lead to a burst of star formation.
The similarities in the global properties of \HII\ galaxies and LSB dwarfs
({\it e.g.} \HI\ mass, \HI\ line width) are consistent with this, although
larger and more complete samples (especially for the LSB dwarfs) should
be studied to confirm this idea.

Taylor \etal \markcite{TBPS}(1994) suggested a possible physical mechanism
by which companions could trigger bursts of star formation in \HII\ galaxies.
It was proposed that the interaction drove gas to the center of the galaxy,
where it accumulated until some physical condition (such as rising above a
threshold surface density) was met, and a burst of star formation began.
Radial inflows of gas have been seen to occur in numerical models of
interacting galaxies (\eg Noguchi \markcite{No87}\markcite{No88}1987, 1988;
Hernquist \markcite{H89}1989; Mihos \& Hernquist \markcite{MH94}1994).
Many, but not all, of the \HII\ galaxies have their star formation bursts
occuring at the center of the galaxy (Salzer \etal \markcite{SMB2} \etal
(1989b), therefore some alternate mechanism is needed to explain the 
off--center systems.  

Under the Taylor \etal \markcite{TBPS}(1994) scenario, LSB dwarfs should
have \HI\ distributions where the surface central density of \HI\ is lower 
than in \HII\ galaxies.  It is known that the total surface density of
LSB galaxies in general is less than for HSB systems ({\it e.g.} McGaugh
\markcite{McG}1996).  To explore any relationship between the \HII\ 
galaxies and the LSB dwarfs will require high resolution observations to 
compare their \HI\ distributions, kinematics and rotation curves.  Some 
of the necessary data have already been obtained.  LSB galaxies are 
usually dominated by their dark matter components (Zwaan \etal \markcite{ZVDM}
1995), a situation which would tend to make them stable against global 
perturbations 
which could trigger star formation bursts.  This stability could play a role
in determining which LSB dwarfs become \HII\ galaxies and which do not,
although some \HII\ galaxies are also known to be dark matter dominated
({\it e.g.} NGC~2915; Meurer \etal \markcite{M96} 1996), so clearly dark 
matter cannot always provide enough stability to prevent a star formation
burst.  There is a lack in the literature of computer simulations of dwarf 
galaxy encounters that might shed some light on the role of dark matter and
what sorts of dwarf--dwarf encounters are necessary dynamically to lead to
the kind of interaction described by Taylor \etal \markcite{TBPS}(1994).

The result of this paper suggests that interactions play an important part 
in triggering bursts of star formation in \HII\ galaxies.  However, 
interactions cannot be the only mechanism at work because in the sample of 
TBGS \markcite{TBGS} 9/21 \HII\ galaxies had no detected companions.  In 
section 3.2.1 I showed that there were likely companions missing from the 
TBGS survey at large radii from the \HII\ galaxies, but I could not determine
the extent of this incompleteness.  It is possible that some \HII\ galaxies 
have no companions, in which case some other mechanism would be responsible
for the bursts of star formation.  It is also apparent that the mere presence
of a companion to interact with is not sufficient to trigger a burst of
star formation because 4/17 LSB dwarfs have companions but no bursts.  This
could be explained if LSB dwarfs were {\it more} dark matter dominated
than \HII\ galaxies, but questions of how much dark matter is needed to
stabilize against interactions of any given strength are best sorted out 
by numerical simulations which explore the parameter spaces relevant to 
encounters among dwarf galaxies.

\subsection{Caveats}

There are two items which could cloud the relatively straightforward 
interpretation discussed above.  The first concerns two of the
galaxies in the \HII\ galaxy sample.  As discussed by TBGS\markcite{TBGS},
two of the \HII\ galaxies are in fact small spirals (UM477 and UM499) and
not dwarf galaxies.  They were retained in the analysis because they 
have bursts of star formation in their nuclei (Salzer \etal \markcite{SMB2}
1989{\it b}), and the goal of the study is to investigate a possible 
connection between bursts and galaxy interactions.  Dwarf galaxies are
excellent for such a study because they are less complex systems than spirals 
and do not have spiral density waves which can also serve as a star formation 
trigger.  Thus by including two spirals in the sample, I am possibly adding
two galaxies to the sample which may not belong there.  I feel justified
in including the two spirals because their bursts are at the galaxy centers,
where the rotation curve shows solid body rotation.  Thus the centers of 
these galaxies have physical conditions similar to the other \HII\ galaxies
(Taylor \etal \markcite{TBS}(1993), TBGS).
However, I note that if the two spiral galaxies are excluded from the
sample, the companion rate becomes 10/19 = 0.53, with a lower limit of
0.32.  This is compared to 12/21 = 0.57, with a lower limit of 0.37.  Thus
the companion rate decreases, but the lower limit is still above the 
companion rate measured for the LSB dwarfs (0.24).

The second item concerns the sample of LSB dwarfs.  Because of the difficulties
inherent in identifying LSB galaxies (Disney \& Phillipps \markcite{DP83} 1983)
no complete samples exist.  This is true of the source lists from which TTBS
drew their sample of LSB dwarfs ({\it e.g.,} Schombert \etal 
\markcite{SBSM}1992).  Thus
it is difficult to quantitatively characterize the LSB dwarf sample.  One 
example of how this can lead to trouble is the case of LSB dwarf L1-137 and
its HI--rich companion, L1-137A.  Inspection of digitized POSS images found 
that the 
companion has a faint, low surface brightness optical counterpart, one which 
was not included in the original source list of LSB dwarfs.  If this galaxy 
had been included in the original sample of LSB dwarfs, then both L1-137 and 
L1-137A would be considered as companions to LSB dwarfs (in a binary pair) and 
the companion rate would be 5/18 = 0.28, instead of 4/17 = 0.24.  Because
the LSB sample is incomplete I cannot estimate how many more binary LSB 
dwarf systems there are where not even one member made the TTBS sample. The 
L1-137
system is the only such case found by TTBS, and would make a small change in
the companion rate, but it does amply illustrate the need for both larger
samples and more complete samples (or at least samples with better quantified 
incompleteness).  The best solution to this problem would be a highly 
sensitive \HI\ mapping survey over a relatively large ($\sim$ 300 degrees$^2$).
Such a survey would have similar sensitivities to both LSB dwarfs and \HII\
galaxies, which could be distinguished from each other in follow--up optical
work. 

\section{Conclusions}

I have used \HI\ surveys of \HII\ galaxies (Taylor \etal \markcite{TBGS}1995; 
Taylor \etal \markcite{TBGS2}1996) and LSB dwarf galaxies (Taylor
\etal \markcite{TTBS}1996b) to study the relationship between galaxy
environment and bursts of star formation in dwarf galaxies.  
The primary result of this work is:

\begin{enumerate}

\item{} \HII\ galaxies have a rate of companion occurrence more than twice
as high as do LSB dwarfs (0.57 compared to 0.24).  The lower limit on the
companion frequency for \HII\ galaxies is 0.37, still higher than the
observed rate for LSB dwarfs.  Because of the incompleteness in projected
separation from the parent galaxy that affects the companion samples,
upper limits are impossible to determine.  However, I have shown that
this incompleteness will affect both samples in the same way, and
will not cause a relative bias between them. Thus I conclude that
this difference in the companion frequency is genuine.

\vspace{.25in}

\hspace{-6ex} This is supported by the following results:

\vspace{.25in}

\item The distribution of companions to \HII\ galaxies detected by
Taylor \etal \markcite{TBGS} \markcite{TBGS2}(1995; 1996a) is likely 
complete in velocity separation 
from the \HII\ galaxy.  It is unlikely that any physically associated 
companions were missed because they were outside the velocity coverage
of the observations.  The distribution of companions to LSB dwarfs 
detected by Taylor \etal (1996b) is not well determined due to the small
number of objects detected.  However, because the LSB dwarfs have 
dynamical masses similar to the \HII\ galaxies, and were observed in exactly
the same fashion as the \HII\ galaxies, their distribution is likely 
complete as well.

\item The distribution of companions to \HII\ galaxies is likely {\it not}
complete in projected linear separation.  This is because the fixed angular
size of the VLA primary beam images larger areas as the distance of the target
galaxy increases.  Taylor \etal \markcite{TBGS}\markcite{TBGS2}(1995; 1996a)
detected companions around the most distant \HII\ galaxies that were so far
from their parent galaxies that they would not have been seen around the
closest \HII\ galaxies.  Without knowing in an unbiased fashion how the 
companions are distributed with respect to their parent galaxies, the 
incompleteness due to this effect cannot be estimated.  Similarly the 
distribution of companions to LSB dwarfs is likely not complete in projected 
linear separation.

\item The lower limit \HI\ sensitivity of the two surveys is approximately
a few $\times~10^7$ M$\sun$ over most of the distance range of the two samples.
The samples of companions are not likely missing objects down to this level
of sensitivity, though there could be objects less massive than this that
remain undetected.

\item The large scale environments (out to separations of 2.5 Mpc) of 
the \HII\ galaxies and the LSB dwarfs are very similar, based on searches
for nearby neighbors and groups in published galaxy catalogs.  Within the
errors, the distribution of neighbor separations for \HII\ galaxies and
LSB dwarfs is the same.  Also, there is no significant difference in the
fraction of \HII\ galaxies and LSB dwarfs which belong to galaxy groups.

\end{enumerate}

The high rate of companion occurrence for \HII\ galaxies relative to LSB 
dwarfs supports the hypothesis that interactions can trigger the bursts 
of star formation seen in \HII\ galaxies.  This implies that \HII\
galaxies may be LSB dwarfs in which a burst of star formation has
occured, but more evidence is needed to confirm this idea.  The 
data in the two surveys do not suggest by what mechanism star formation
bursts can be triggered, although Taylor \etal \markcite{TBPS}(1994) do 
offer one possibility.

\acknowledgements

I thank Evan Skillman and Elias Brinks, my thesis advisors, for their many 
contributions, both direct and indirect, that led to this paper.  I am
also grateful to John Salzer for helpful conversations and to Christine
Wilson and Ralph Pudritz for comments on an early version of the manuscript.
I thank Greg Bothun and the anonymous referee for useful comments which 
improved the paper.
This research has made use of the NASA/IPAC Extragalactic Database (NED)
which is operated by the Jet Propulsion Laboratory, California Institute
of Technology, under contract with the National Aeronautics and Space
Administration.  This research has made use of NASA's Astrophysics Data
System Catalog Service.

\newpage

\figcaption[] {Upper: A histogram showing the distribution of recession
velocities for the sample of \HII\ galaxies.  Lower: A histogram showing 
the distribution of recession velocities for the sample of LSB dwarfs.
\label{fig1}}

\figcaption[] {Upper: The bold face histogram shows the distribution of velocity
separations of the companions from their parent \HII\ galaxies.  The shaded
histogram is a sample made by randomly selecting \HII\ galaxies and companions
and computing the velocity difference.
Lower: The bold face histogram shows the distribution of velocity
separations of the companions from their parent LSB dwarf galaxies.  The shaded
histogram is a sample made by randomly selecting LSB dwarfs and companions
and computing the velocity difference.
\label{fig2}}

\figcaption[] {Upper: A histogram of the distribution of projected linear 
separations of the companions from their parent \HII\ galaxies.  The vertical 
lines shows the edge of the 30$\arcmin$ VLA primary beam at the distance of 
the closest \HII\ galaxy.
Lower: A histogram of the distribution of projected linear separations
of the companions from their parent LSB dwarf galaxies.  The vertical lines 
shows the edge of the 30$\arcmin$ VLA primary beam at the distance of the 
closest LSB dwarf.
\label{fig3}}

\figcaption[] {A histogram of the distribution of angular separations
of the companions from their parent \HII\ galaxies.  The dashed line
shows the radius FWHM of the VLA primary beam. 
\label{fig4}}

\figcaption[] {Upper: A plot of \HI\ mass versus $v_{corr}$ for the \HII\ 
galaxies (open triangles) and their companions (y--shaped symbols).  The 
curve shows the \HI\ sensitivity as a function of distance for a typical 
observation.  Lower: A plot of \HI\ mass versus $v_{corr}$ for the LSB dwarfs
(open circles) and their companions (stars).  The curve shows the \HI\ 
sensitivity as a function of distance for a typical observation.
\label{fig5}}

\figcaption[] {The mean number of neighbors in 0.5 Mpc wide bins versus 
separation from the sample galaxies.  The open symbols show the mean counts 
in each bin, while solid symbols show cumulative mean counts.  The dashed 
vertical line shows the maximum separation from a galaxy observable with 
one pointing of the VLA.
\label{fig6}}

\vfill \eject
\begin{figure}
\psfig{figure=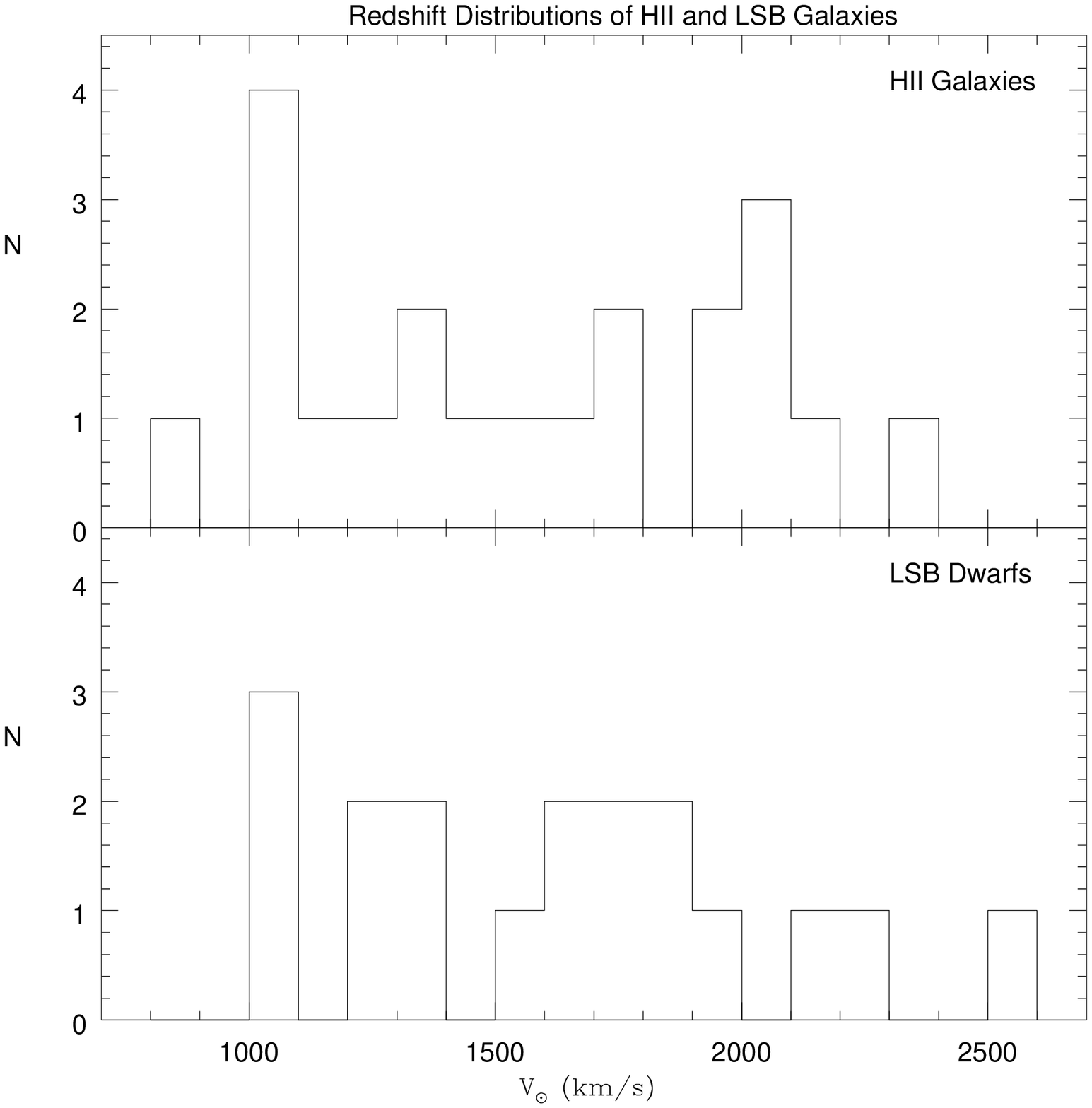,bbllx=000bp,bblly=250bp,bburx=370bp,bbury=700bp,height=5.5in}
\vspace{2in}{Figure 1}
\end{figure}

\vfill \eject
\begin{figure}
\psfig{figure=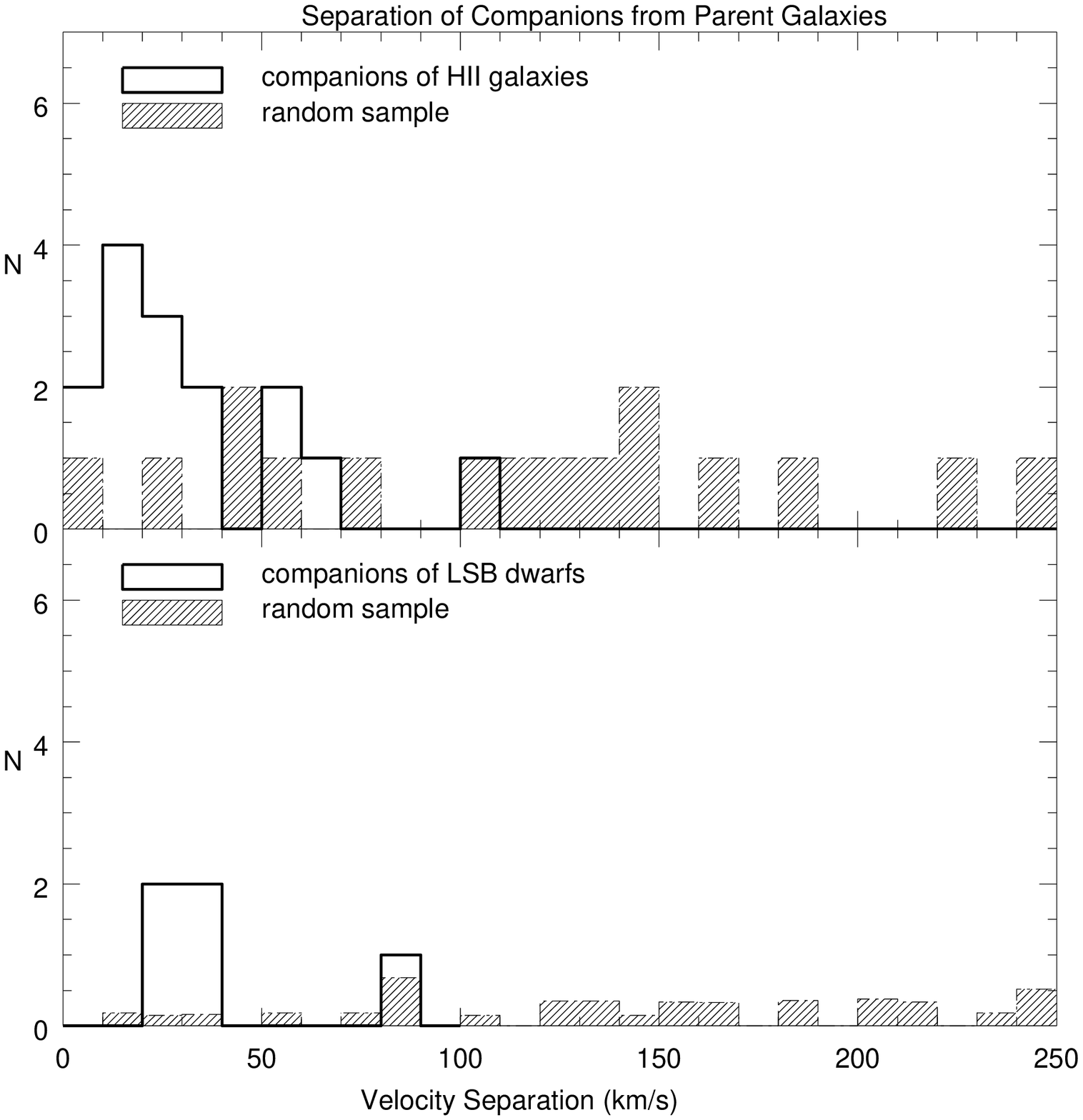,bbllx=000bp,bblly=250bp,bburx=370bp,bbury=700bp,height=5.5in}
\vspace{2in}{Figure 2}
\end{figure}

\vfill \eject
\begin{figure}
\psfig{figure=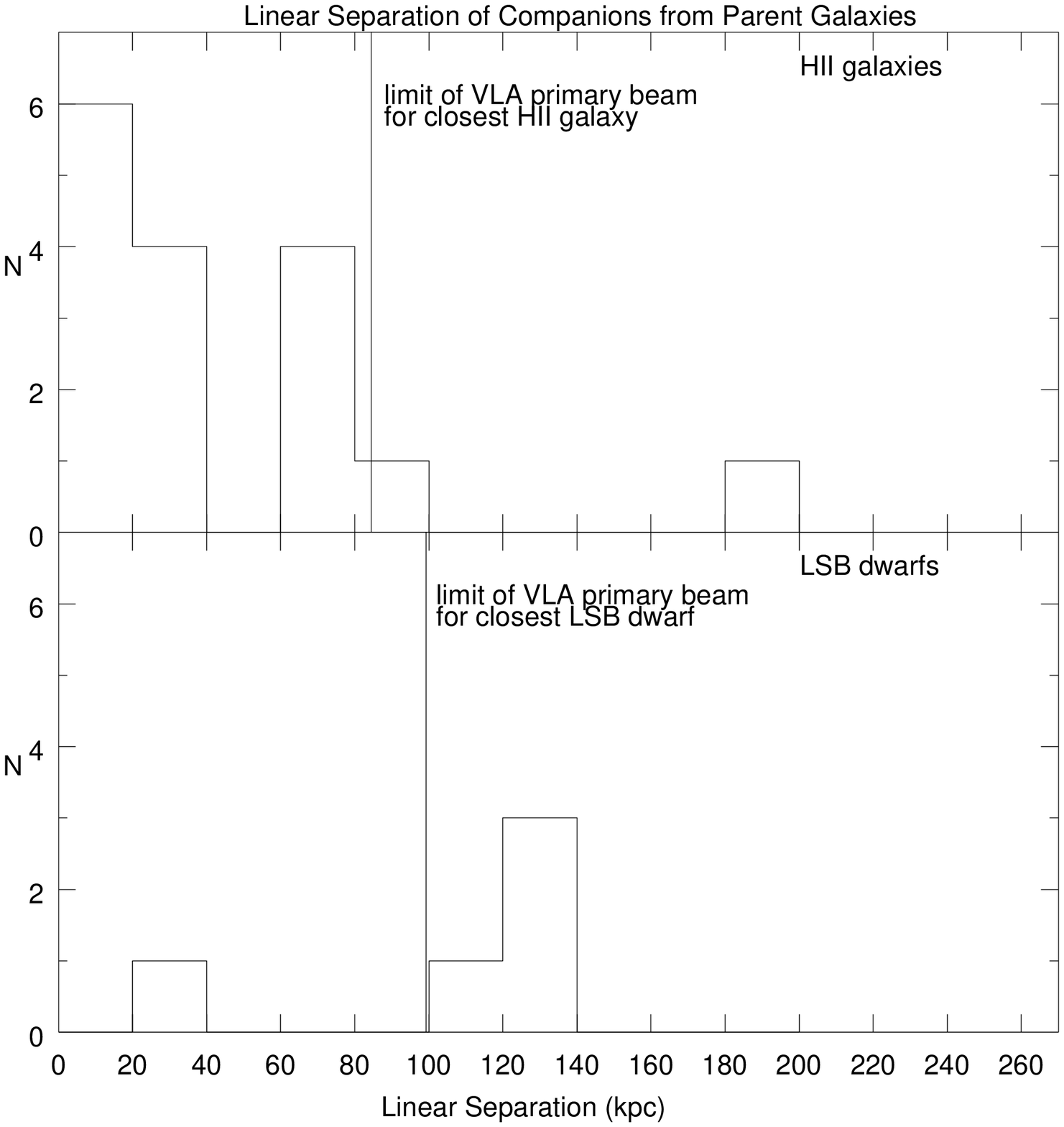,bbllx=000bp,bblly=250bp,bburx=370bp,bbury=700bp,height=5.5in}
\vspace{2in}{Figure 3}
\end{figure}

\vfill \eject
\begin{figure}
\psfig{figure=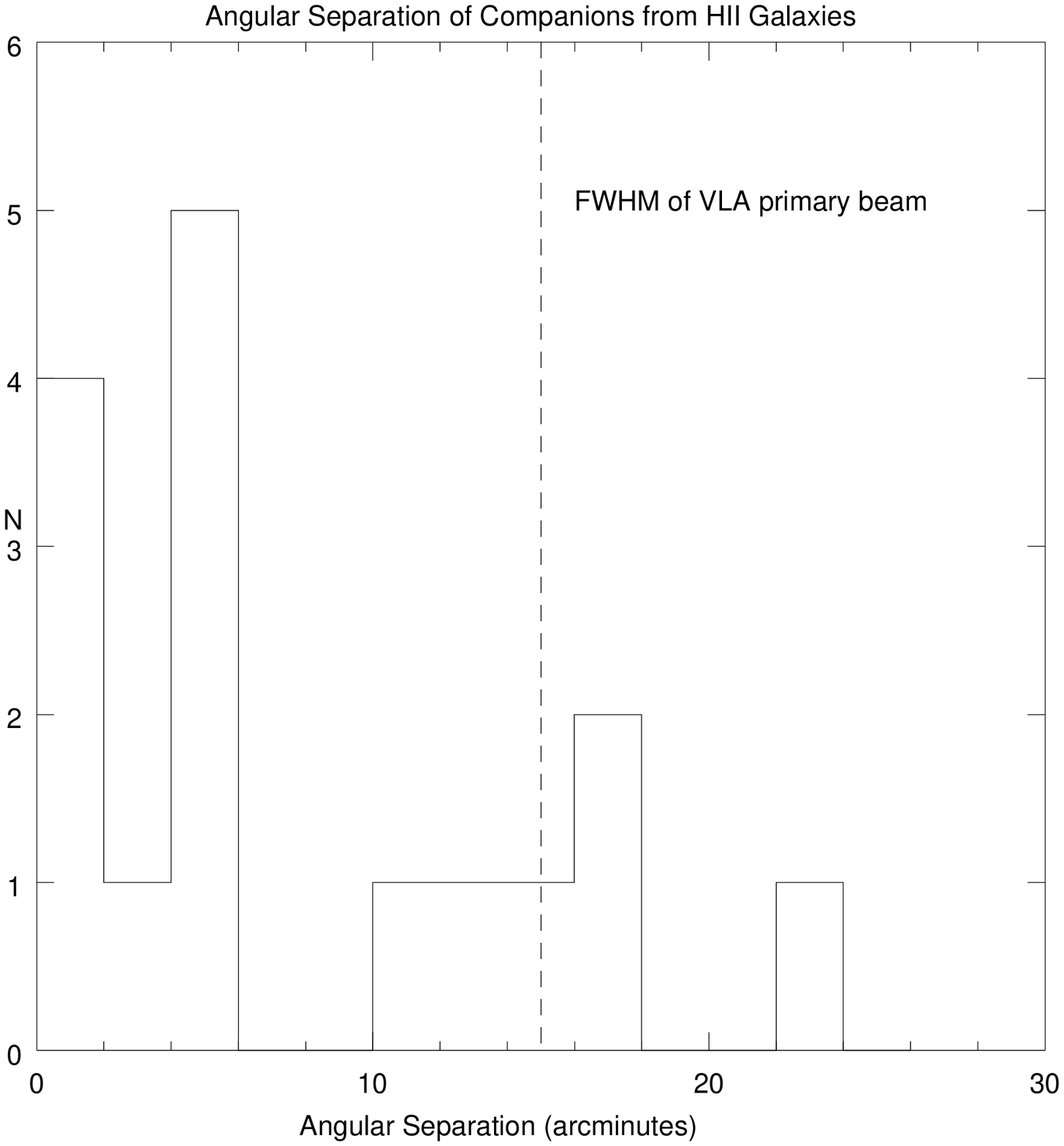,bbllx=000bp,bblly=250bp,bburx=370bp,bbury=700bp,height=5.5in}
\vspace{2in}{Figure 4}
\end{figure}

\vfill \eject
\begin{figure}
\psfig{figure=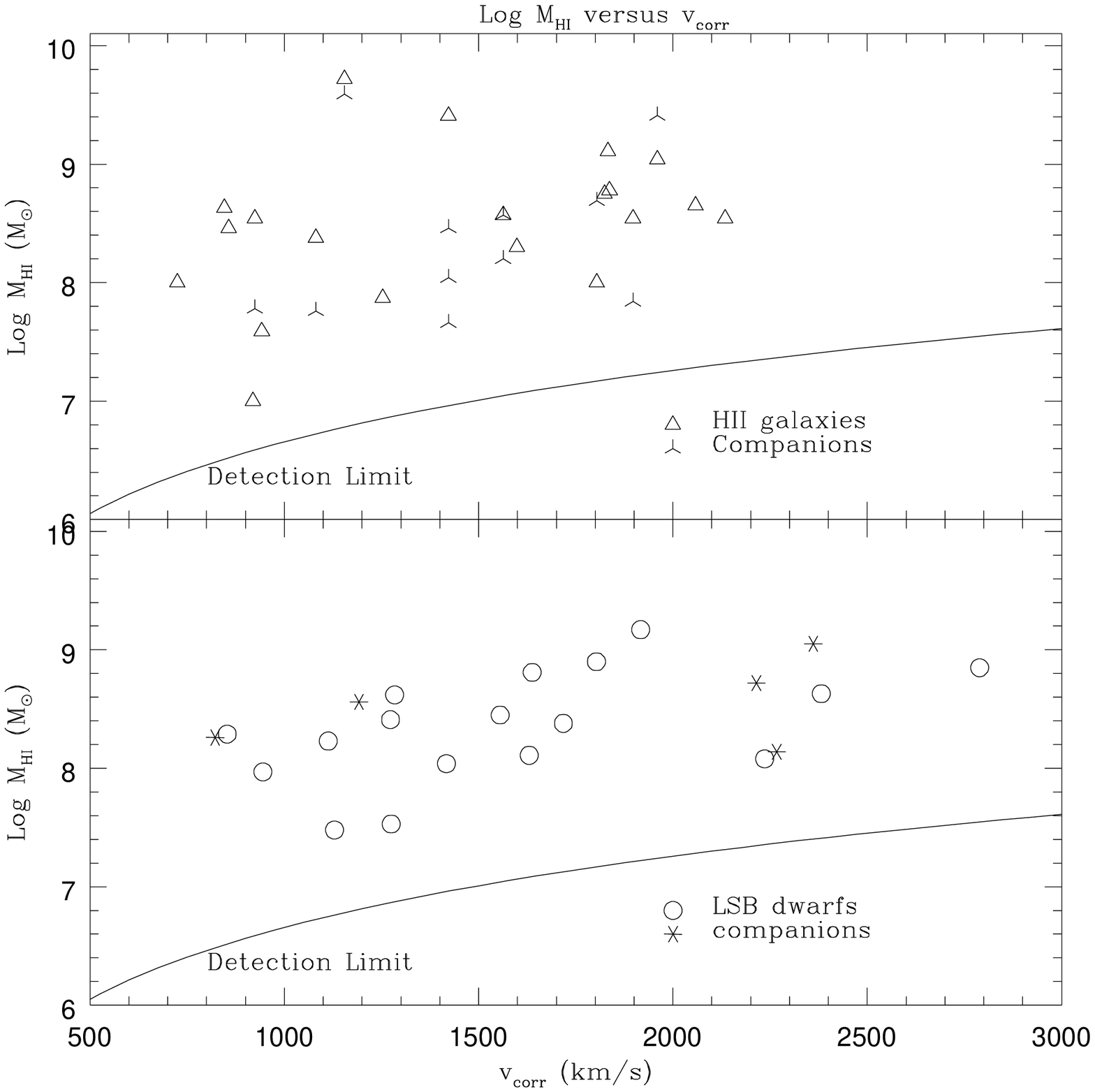,bbllx=000bp,bblly=250bp,bburx=370bp,bbury=700bp,height=5.5in}
\vspace{2in}{Figure 5}
\end{figure}

\vfill \eject
\begin{figure}
\psfig{figure=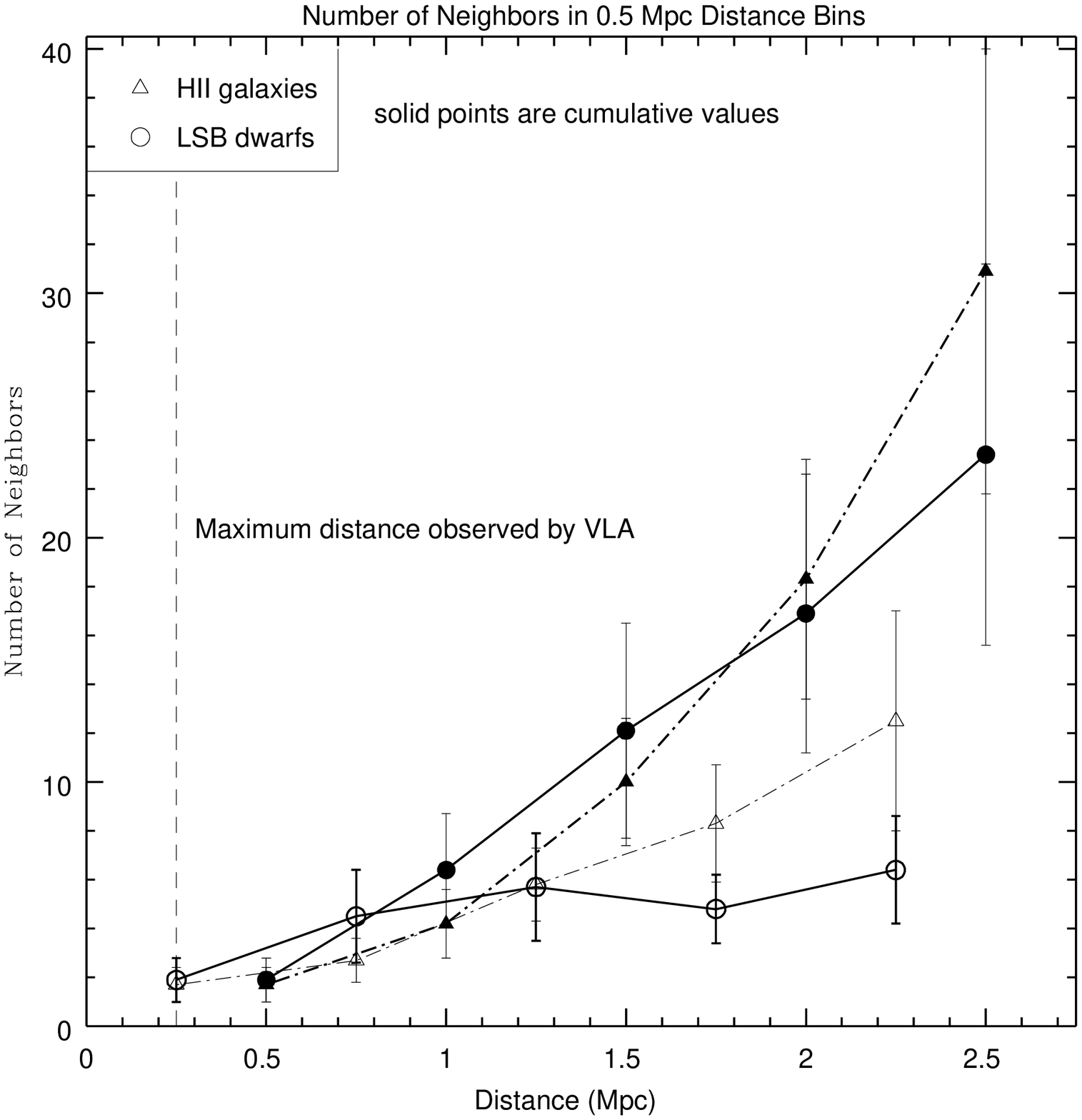,bbllx=000bp,bblly=250bp,bburx=370bp,bbury=700bp,height=5.5in}
\vspace{2in}{Figure 6}
\end{figure}

\end{document}